\documentstyle[prl,aps,multicol]{revtex}

\begin{document}

\title{Variational methods, multiprecision and nonrelativistic
energies}

\author{V.I.~Korobov}
\address{Joint Institute for Nuclear Research,\\
        141980, Dubna, Russia}

\maketitle

\begin{abstract}
It is known that the variational methods are the most powerful tool
for studying the Coulomb three--body bound state problem. However,
they often suffer from loss of stability when the number of basis
functions increases. This problem can be cured by applying the
multiprecision package designed by D.H.~Bailey. We consider the
variational basis functions of the type $\exp(-\alpha_n r_1-
\beta_n r_2-\gamma_n r_{12})$ with complex exponents. The method
yields the best available energies for the ground states of the
helium atom and the positive hydrogen ion as well as many other
known atomic and molecular systems.
\end{abstract}
\pacs{31.15.Ar, 31.15.Pf}

\begin{multicols}{2}

{\bf1.} The development of the variational method for the Coulomb
bound state problem can be traced using as an example the ground
state of the helium atom. In early days when computers were big and
very expensive the search proceeded mainly in the direction of
making expansion of the variational wave function as compact as
possible (in a sense of number of variational parameters and/or
basis sets). At first, the explicitly correlated basis were
introduced \cite{Hylleraas,Pekeris1} now called as the Hylleraas
basis
\[ \begin{array}{@{}l}\displaystyle
\psi({\bf r}_1,{\bf r}_2) =
e^{-\frac{1}{2}s}\sum c_{lmn}s^lu^mt^m, \\[2mm]
\hspace{5mm} s=r_1+r_2,\quad u=r_{12},\quad t=-r_1+r_2,
\end{array} \]
then it became clear that at least for the ground state of the
helium atom it is essential to incorporate into the wave function
such peculiarity as the logarithmic behaviour of the type $~R\ln
R~$ at $R=(r_1^2+r_2^2)^{\frac{1}{2}}\to0$, first analytically
derived independently by Bartlett and Fock \cite{Fock}. In 1966,
Frankowski and Pekeris (see Table \ref{II}) introduced the compact
representation \cite{Pekeris} of the form
\[ \psi({\bf r}_1,{\bf r}_2) =
e^{-\kappa s}\sum c_{lmnij}s^lu^mt^{2m}(s^2+t^2)^{i/2}(\ln{s})^j,
\]
and later, in 1984, Freund and co-workers \cite{Morgan} reported
even more compact expansion of the same form. Inclusion of the
logarithmic term into the variational wave function brought
substantial improvement of nonrelativistic energies for the two
electron atoms. In 1994, Thakkar and Koga \cite{Koga} have found a
compact expansion without logarithms which uses powers that are not
integers nor even half integers. As far as we know none of these
compact expansions has been used for analytical evaluation of
matrix elements of the Breit interaction.

With advance of computer power basis sets became simplified
that allowed for calculation of numerous matrix elements required
for relativistic and QED corrections. The efforts were concentrated
on a choice of a strategy that defines a sequence of basis
functions genereated. In \cite{Yan} the double basis set method
with generalyzed Hylleraas basis functions
\[ \begin{array}{@{}l}\displaystyle
\psi({\bf r}_1,{\bf r}_2) = \sum c_{ijk}r_1^ir_2^jr_{12}^k
e^{-\alpha r_1-\beta r_2}\\[2mm]
\hspace{15mm}\displaystyle+\sum \overline{c}_{ijk}r_1^ir_2^jr_{12}^k
e^{-\overline{\alpha}r_1-\overline{\beta}r_2}
\end{array} \]
were used. This double basis set technique along with full
optimization of nonlinear parameters at each basis set yield
substantial progress in accuracy. However, the main factor that
hinder further advance become the numerical instability due to
almost linear dependence of the basis set at large $N$.

The work of Goldman \cite{Goldman} is a bit apart of the main
path. It recovers the idea of Pekeris \cite{Pekeris1} to use
uncoupled coordinates and orthogonal Laguerre and Jacoby
polynomials as basis functions.

The method expounded in our work is a continuation of efforts by
Drake and Yan to utilize as much simple basis functions (geminals)
as possible.

{\bf2.} Expansion we want to consider here is very similar to the
generalized Hylleraas basis set, but instead of using the
polynomials over Hylleraas variables we generate nonlinear
parameters in the exponents in a quasi-random manner,
\begin{equation}\label{anzatz}
r_1^{l_i}r_2^{m_i}r_{12}^{n_i}
e^{-\alpha r_1-\beta r_2-\gamma r_{12}}
\Longrightarrow e^{-\alpha_i r_1-\beta_i r_2-\gamma_i r_{12}}.
\end{equation}
This method has been successfully used in calculations
\cite{Monkhorst,Efros} previously. Obviously, the matrix elements
can be evaluated in the same way as for the generalized Hylleraas
basis set (\ref{anzatz}). Moreover, if one replaces real exponents
by complex exponents the integrals will remain exactly the same as
for the real case. In its strategy the method is very close to the
SVM method by Varga, Suzuki \cite{SVM}, where gaussians are
exploited instead.

In a formal way, a variational wave function is expanded in a form
\[ \begin{array}{@{}l}
\displaystyle
\psi_0 = \sum_{i=1}^{\infty}
\Big\{U_i\,{\sl Re}
\bigl[\exp{(-\alpha_i r_1-\beta_i r_2-\gamma_i r_{12})}\bigr]
\\[2mm]\displaystyle\hspace{5mm}
+W_i\,{\sl Im}
\bigl[\exp{(-\alpha_i r_1-\beta_i r_2-\gamma_i r_{12})}\bigr]
\Big\}{\cal Y}_{l_1l_2}^{LM}(\hat{\bf r}_1,\hat{\bf r}_2).
\end{array} \]
Here $\alpha_i$, $\beta_i$ and $\gamma_i$ are complex parameters
generated in a quasi-random manner \cite{Frol_mol,KorNew98}:
\[ \begin{array}{l}\displaystyle
\alpha_i = \left\lfloor\frac{1}{2}i(i+1)\sqrt{p_\alpha}
\right\rfloor[(A_2-A_1)+A_1]+\\[3mm]\displaystyle\hspace{5mm}
+i\left\{\left\lfloor\frac{1}{2}i(i+1)\sqrt{q_\alpha}
\right\rfloor[(A'_2-A'_1)+A'_1]\right\},
\end{array}
\]
$\lfloor x\rfloor$ designates the fractional part of $x$,
$p_\alpha$ and $q_\alpha$ are some prime numbers, $[A_1,A_2]$ and
$[A'_1,A'_2]$ are real variational intervals which need to be
optimized. Parameters $\beta_i$ and $\gamma_i$ are obtained in a
similar way.

An important feature of the method is that it demonstrates a very
fast convergence. The general rule which can be inferred
experimentally from the use of the method is that increasing of the
basis by about 200 functions yields about one additional digit in
the variational energy. The minor deficiency is that the basis
quickly degenerates when $N$ increases. Already for moderate
$N\sim250-400$ a quadruple precision is required.

Multiprecision package of Fortran routines MPFUN has been designed
by David H.~Bailey \cite{Bailey} for computations with floating
point numbers of an arbitrary length. Usually it is necessary to
make significant changes into Fortran source code in case if
Fortran-77 language is used. Fortunately, the author of MPFUN
package has developed a translator program that facilitate
converting the programs to multiprecision drastically. In general,
two directives incorporated as comments in a source code are
required per one routine. For example a source code for the
considered variational method has been transformed to
multiprecision version within two hours of manual work. Eventually
a code we've gotten has been tested on a personal computer with the
Celeron 500 MHz processor. For one run with the basis of $N=1400$
functions and 40 decimal digits it requires about 3 hours.

For users of Fortran--90 no preprocessor is needed due to new
advanced features of Fortran--90, such as derived data types and
operator extensions.

\noindent\begin{minipage}{0.48\textwidth}
\begin{table}
\begin{center}
\begin{tabular}{c@{\hspace{15mm}}l}
$N$ & \hfil$E$ (a.u.) \\ \hline
1400 & $-$2.90372437703411959629   \\
1600 & $-$2.903724377034119597843  \\
1800 & $-$2.9037243770341195981964 \\
2000 & $-$2.9037243770341195982713 \\
2200 & $-$2.9037243770341195982955 \\ \hline
extrapolation & $-$2.903724377034119598306(10)\\
\end{tabular}
\end{center}
\caption{Variational energy (in a.u.) of the helium ground state
as a function of $N$, the number of basis functions.}
\label{I}
\end{table}
\end{minipage}

In our calculations for the helium ground state four basis sets
with independently optimized nonlinear parameters were used. These
sets were built up like a pine tree. The first layer was tuned to
approximate the general behaviour of the solution at intermediate
and large $r_1$ and $r_2$. The second layer was chosen to be
flexible in a smaller region of $r_1$ and $r_2$ and so forth. A
detailed optimization was performed for the sets with total
$N=1400$ and $N=1600$. Quadruple precision was not sufficient at
these $N$ and we used the multiprecision version of the program
with 40 significant decimal digits. Further calculations with
$N=1800-2200$ were performed with 48 significant digits and only
partial optimization of the parameters of the last layer
(corresponding to the region where the logarithmic behaviour is the
most essential) was done. Some optimization of a distribution of
$n_i$ between the layers ($N=n_1+n_2+n_3+n_4$) was carried out as
well.

As can be seen from the Table \ref{II} the present result extends
the accuracy of the nonrelativistic ground state energy for the
helium atom by as much as 3 decimal digits.

\noindent\begin{minipage}{0.48\textwidth}
\begin{table}
\begin{center}
\begin{tabular}{l@{\hspace{3mm}}rl}
                            & $N$~& \hfil$E$ (a.u.) \\ \hline
Frankowski and              & 246 & $-$2.9037243770326        \\
~~~~~~Pekeris \cite{Pekeris}&     &                           \\
Freund, Huxtable,           & 230 & $-$2.9037243770340        \\
~~~~~~and Morgan III \cite{Morgan}&&                          \\
Thakkar and Koga \cite{Koga}& 308 & $-$2.9037243770341144     \\
Drake and Yan \cite{Yan}    & 1262& $-$2.90372437703411948    \\
Goldman \cite{Goldman}      & 8066& $-$2.903724377034119594   \\
This work                   & 2200& $-$2.903724377034119598296\\
\end{tabular}
\end{center}
\caption{Comparison of the ground state energy of the helium atom
obtained in this work with other theoretical calculations.}
\label{II}
\end{table}
\end{minipage}

Second case is the hydrogen molecular ion ground state that
represent an other limit of mass distribution of constituents with
one light and two heavy particles. For this case it is especially
essential that we introduce complex exponents, because it is the
most natural way to suit the oscillatory behaviour of the
vibrational motion in the wave function. In this case (see Table
\ref{III}) again 40 decimal digits have been used for $N=1400-1800$
and 48 decimal digits for large $N$ to provide the numerical
stability of the calculations. Table \ref{IV} demonstrates progress
in obtaining variational nonrelativistic energy for this state. The
accuracy is extended by as much as 4 additional digits.

\noindent\begin{minipage}{0.48\textwidth}
\begin{table}
\begin{center}
\begin{tabular}{c@{\hspace{15mm}}l}
$N$ & \hfil$E$ (a.u.) \\ \hline
1400 & $-$0.597139063123404975  \\
1600 & $-$0.597139063123405047  \\
1800 & $-$0.5971390631234050655 \\
2000 & $-$0.5971390631234050710 \\
2200 & $-$0.5971390631234050740 \\ \hline
extrapolation & $-$0.597139063123405076(2) \\
\end{tabular}
\end{center}
\caption{Variational energy (in a.u.) of the positive hydrogen ion
ground state as a function of $N$, the number of basis functions.}
\label{III}
\end{table}
\end{minipage}

\noindent\begin{minipage}{0.48\textwidth}
\begin{table}
\begin{center}
\begin{tabular}{l@{\hspace{5mm}}rl}
                            &$N$~~& \hfil$E$ (a.u.) \\ \hline
Gr\'emaud, Delande          &31746& $-$0.597139063123        \\
~~~~~~and Billy \cite{Gremaud}&   &                          \\
Rebane and Filinsky \cite{Rebane}&& $-$0.59713906312340      \\
Moss \cite{Moss}            &     & $-$0.5971390631234       \\
This work                   & 2200& $-$0.597139063123405074  \\
\end{tabular}
\end{center}
\caption{Comparison of the ground state energy of the positive
hydrogen molecular ion obtained in this work with other theoretical
calculations. $m_p=1836.152701 m_e$.}
\label{IV}
\end{table}
\end{minipage}

In Table \ref{V} the other examples are summarized. A negative
positronium ion demonstrates a limit of three particles of equal
masses. The second and third cases are applications of the method
to the states with nonzero angular momentum. The last example in
this Table is of special proud. That is the last vibrational state
in a series of $S$-states of the hydrogen molecular cation,
and that is the first variational confirmation of the existence of
this state (the binding energy corresponding to the cited value is
0.74421(2) cm$^{-1}$). The accuracy of the artificial channels
scattering method \cite{Moss2} is presumably better, however, wave
functions are not forthcoming with this method that makes difficult
calculation of physical properties of the state other than energy.

\noindent\begin{minipage}{0.48\textwidth}
\begin{table}[t]
\begin{center}
\begin{tabular}{l@{\hspace{1mm}}rl}
~~~~~system        &             & \hfil$E$                   \\ \hline
$^{~}e^-e^-e^+$    & This work   & $-$0.2620050702329801077(3)\\
                   &\cite{Frolov}& $-$0.262005070232976       \\[2mm]
$^{~}$He$(2^3P)$   & This work   & $-$2.13316419077928310(2) \\
                   & \cite{Yan2} & $-$2.13316419077927(1)    \\[2mm]
$^4$He$^+\bar{p}\,(L\!=\!35,v\!=\!0)$
                   &This work    & $-$2.98402095449725(1) \\
                   &\cite{Kino}  & $-$2.98402094          \\[2mm]
$^{~}$H$_2^+\,(L\!=\!0,v\!=\!19)$
                   & This work   & $-$0.4997312306        \\
                   & \cite{Moss2}& $-$0.49973123063       \\
\end{tabular}
\end{center}
\caption{Other examples of three--body calculations. ($L$ is the
total angular momentum, $v$ is the vibrational quantum number.)}
\label{V}
\end{table}
\end{minipage}

{3.} One may say that this high accuracy is redundant and has no
physical meaning. But obviously, it shows the power of modern
computers and theirs ability to solve the quantum three--body
problem to any required accuracy. On the other hand, uncertainty in
the variational wave function approximately as much as the
square root of the uncertainty in the variational energy and is
about $10^{-9}-10^{-10}$. This accuracy does not look redundant.
These results prove that the nonrelativistic bound state
three--body problem is now satisfactorily solved and the main
efforts should be addressed to relativistic and QED effects.

The other advantage of the method is the simplicity of the basis
functions that allows for evaluate analytically relativistic matrix
elements of the Breit Hamiltonian. It is possible as well to
evaluate analytically the vacuum polarization term (Uehling
potential) \cite{Petelenz} and to build up an effective numerical
scheme for the one--loop self--energy corrections \cite{BetheLog}.
These features make the considered variational method to be highly
powerful universal tool for studying the three--body problem.

This work has been partially supported by INTAS Grant No. 97-11032,
which is gratefully acknowledged.


\end{multicols}

\end{document}